\documentclass[preprint,showpacs,preprintnumbers,amsmath,amssymb]{revtex4}
\usepackage{graphicx}

\renewcommand{\epsilon}{\varepsilon}
\newcommand{\diff}{\mathrm d}

\begin{document}
\title{
Mapping functions and critical behavior of percolation on rectangular domains}
\author{Hiroshi Watanabe$^{1}\footnote{E-mail: hwatanabe@is.nagoya-u.ac.jp.}$
and Chin-Kun Hu$^{2,3}$\footnote{E-mail: huck@phys.sinica.edu.tw.}}

\affiliation{$^1$Department of Complex
Systems Science, Graduate School of Information Science, Nagoya
University, Furouchou, Chikusa-ku, Nagoya 464-8601, Japan}

\affiliation{$^2$Institute of Physics, Academia
Sinica, Nankang, Taipei 11529, Taiwan}

\affiliation{$^3$Center for Nonlinear and Complex Systems and
Department of Physics, Chung-Yuan Christian University, Chungli
32023, Taiwan}

\date{\today}

\begin{abstract}
The existence probability $E_p$ and the percolation probability
$P$ of the bond percolation on rectangular domains with different
aspect ratios $R$ are studied via the mapping functions between
systems with different aspect ratios.
The superscaling behavior of $E_p$ and $P$ for such systems with
exponents $a$ and $b$, respectively, found by Watanabe, Yukawa,
Ito, and Hu in [Phys. Rev. Lett. \textbf{93}, 190601 (2004)] can
be understood from the lower order approximation of the mapping
functions $f_R$ and $g_R$ for $E_p$ and $P$, respectively; the
exponents $a$ and $b$ can be obtained from numerically determined
mapping functions $f_R$ and $g_R$, respectively.
\end{abstract}

\pacs{02.50.Ng ,05.50.+q, 89.75.Da}

\maketitle


\section{Introduction}

Universality and scaling in critical systems have attracted much
attention in recent decades
\cite{71Gene,fisher70,84prb-pf,privman90}. It has been found that
critical systems can be classified into different universality
classes so that the systems in the same class have the same set of
critical exponents \cite{71Gene}. According to the theory of
finite-size scaling \cite{fisher70,84prb-pf,privman90}, if the
dependence of a physical quantity $M$ of a thermodynamic system on
a parameter $t$, which vanishes at the critical point $t=0$, is of
the form $M(t)\sim t^a$ near the critical point, then for a finite
system of linear dimension $L$, the corresponding quantity
$M(L,t)$ is of the form
\begin{equation}
  M(L,t) \sim L^{-ay_t}F(tL^{y_t}), \label{e0}
\end{equation}
where $y_t$ ($\equiv \nu ^{-1}$) is the thermal scaling power,
$F(x)$  is the scaling function, and $x \equiv tL^{y_t}$ is the
scaling variable. It follows from (\ref{e0}) that the scaled data
$M(L,t)L^{ay_t}$ for different values of $L$ and $t$ are described
by a single finite-size scaling function (FSSF) $F(x)$. Thus it is
important to know general features of the scaling function under
various conditions. On the basis of the renormalization group
arguments, Privman and Fisher \cite{84prb-pf,privman90} proposed
that systems in the same universality classes can have universal
finite-size scaling functions (UFSSFs).


Percolation models
\cite{78jspWuFY,80rppEssamJW,80prb-LCMCRG,90pra-ccp,92hu,92ziff,sa94,rev,ziff96}
are ideal systems for studying universal and scaling behaviors
near the critical point. The system is defined to be percolated
when there is at least one cluster extending from the top to the
bottom of the system. The average fraction of lattice sites in the
percolating clusters is called the percolation probability
\cite{78jspWuFY,80rppEssamJW} (also known as percolation strength)
and will be denoted by $P$ in the present paper. The probability
that at least one percolating cluster exists in the system is
called the existence probability \cite{92hu} or crossing
probability \cite{{ziff96}}, and will be denoted by $E_p$ in the
present paper \cite{Ep-name}.

Using a histogram Monte Carlo method \cite{92hu} and other Monte
Carlo methods, Hu and collaborators found that the FSSFs depend
sensitively on the boundary conditions \cite{94jpa96prl} and
shapes \cite{95jpa} of the lattices and many two-dimensional
percolation models (including bond and site percolation models on
square, honeycomb, and triangular lattices, continuum percolation
of soft disk and hard disks, bond percolation on random lattices,
etc) \cite{perco} can have universal FSSFs (UFSSFs) for their
$E_p$ and $P$, and many  three-dimensional percolation models
(including site percolation on sc, bcc, and fcc lattices and bond
percolation on sc lattice) \cite{perco3d} have universal critical
exponents and UFSSFs for their $E_p$ and $P$.

In the present paper, we will study the bond percolation model on
$L_1 \times L_2$ rectangular lattices (domains)
\cite{Monetti1991,Cardy}, especially the dependence of $E_p$ and
$P$ on the aspect ratio $R$, which is the ratio of the horizontal
length $L_1$ to the vertical length $L_2 \equiv L$ of the system,
i.e., $R=L_1/L_2$. For this system, it is well known that the
critical bond density $\rho_c$ (also called bond occupation
probability) is 1/2, the critical exponent $\beta$ for the
percolation probability $P$ is 5/36, $y_t=1/\nu=3/4$, and the
critical exponent for the existence probability $E_p$ is 0
\cite{sa94,Ep}, which and Eq. (\ref{e0}) imply  that the
dependence of $E_p$ on the the linear dimension $L$, the bond
density $\rho$, and the aspect ratio $R$ can be written as
\begin{equation}
E_p(L,\rho,R) = F_R(\epsilon L^{y_t}), \quad (\epsilon \equiv |\rho-\rho_c|/\rho_c)
\label{eq_ufssf}
\end{equation}
near the critical point $\rho_c$~\cite{84prb-pf,sa94}. The
function $F_R$ is universal for various percolation models
considered in Ref. \cite{perco}. The aspect ratio dependence of
the function $F_R$ is not clear; {\it e.g.}, the systems with
close values of $R$ should have similar forms of $F_R$, however,
Eq.~(\ref{eq_ufssf}) do not tell us how similar they are. While
the value of the existence probability at the critical point
$E_p(\rho_c)$ was obtained by Cardy~\cite{Cardy}, such critical
$E_p$ approaches quickly to $1$ as the aspect ratio is increased,
{\it e.g.}, it is difficult to distinguish the value for $R=16$
from that for $R=32$. Additionally, the region for the rapid
increase of $E_p$  shifts to the smaller value of $\rho$ with
larger $R$ and fixed $L$, which cannot be described
quantitatively. Therefore, we need other kinds of approach to
study the aspect ratio dependence of the UFSSFs.

Recently, we found ``superscaling'' of the existence probability
$E_p$ and the percolation probability $P$ for bond percolation on
rectangular domains with different aspect ratios \cite{04prl}:
$E_p(L,\rho,R) \sim F(\epsilon'L^{y_t}R^a)$ and $P(L,\rho,R) \sim
(L^{y_t}R^b)^{-\beta} F(\epsilon'L^{y_t}R^b)$, with new exponents
$a$ and $b$, where $\epsilon' \equiv (\rho - \rho_c')$ with
$\rho_c'$ being the effective critical point; $a$ and $b$ were
determined by a fitting procedure to be $a=0.14(1)$ and
$b=0.05(1)$, respectively. In the present paper, we define mapping
functions to connect UFSSFs for percolation on domains with
different aspect ratios and find that $a$ and $b$ can be obtained
from numerically determined mapping functions $f_R$ and $g_R$ for
$E_p$ and $P$, respectively~\cite{05prlwh}.





\vskip 2 mm

\section{Mapping function and finite-size scaling}

In order to make the meaning of the mapping function becomes
clear, we first introduce the mapping function for the percolation
model on the square lattices with the same aspect ratio. Consider
two different square systems $A$ and $B$ with the linear sizes
$L_A$ and $L_B (> L_A)$, respectively. The existence probability
of the system $B$ at density (also called occupation probability)
$\rho$ is denoted by $E_p(L_B,\rho)$. The existence probability is
a monotonic increasing function whose value changes from $0$ to
$1$ as the density increases from $0$ to $1$. Therefore, for every
$E_p(L_B,\rho)$ of system $B$, there is a corresponding value
$\rho^{\prime}$ of system $A$ so that the existence probability of
system $A$ at $\rho^{\prime}$, denoted as
$E_p(L_A,\rho^{\prime})$, is equal to $E_p(L_B,\rho)$ and we have
\begin{equation}
E_p(L_B,\rho) = E_p(L_A,\rho^{\prime}) \equiv E_p(L_A,f_Q(\rho)),
\label{eq_map}
\end{equation}
which defines  the renormalization group transformation from
$\rho$ to $\rho'$ in Ref.~\cite{92hu}. The mapping from $\rho$ to
$\rho^{\prime}$ is represented as the mapping function
\begin{equation}
\rho^{\prime}=f_{A \rightarrow B}(\rho)\equiv f_Q(\rho).
\label{rg}
\end{equation}
The mapping function depends only on the ratio of the system size
$Q = L_B/L_A$, and therefore, we denote the mapping function by
$f_Q$. The critical point of the system, $\rho_c$, can be
determined from the fixed point of Eq. (\ref{eq_map}):
$$
E_p(L_B,\rho_c)=E_p(L_A,\rho_c),
$$
which is equivalent to
\begin{equation}
f_Q(\rho_c) = \rho_c. \label{eq_fixedpoint}
\end{equation}
For fixed $Q$, $\rho_c$ of Eq. (\ref{eq_fixedpoint}) approaches
the exact critical point $1/2$ for the bond percolation on the
square lattice as $L_A \to \infty$. Table 1 of the first reference
in \cite{92hu} shows that $\rho_c$ of Eq. (\ref{eq_fixedpoint}) is
already very accurate for $L_B=16-20$ and $L_A=8-16$.

The expansion of the mapping function at the critical point
can be written as
\begin{eqnarray}
\rho' &=& f_Q(\rho) \\
&=& f_Q(\rho_c) + \left. \frac{\diff f_Q}{\diff \rho} \right|_{\rho_c} (\rho-\rho_c)
 + \cdots. \label{eq_size_expansion}
\end{eqnarray}
It is well known that the thermal scaling power $y_t$ can be
obtained from the equation ~\cite{92hu}
\begin{equation}
y_t = \frac{\ln  \left. \displaystyle \frac{\diff f_Q}{\diff
\rho}\right|_{\rho_c}}{\ln Q},
\end{equation}
which implies that the differential coefficient of the first order
in Eq.~(\ref{eq_size_expansion}) is given by
\begin{equation}
\left. \frac{\diff f_Q}{\diff \rho}  \right|_{\rho_c} = Q^{y_t}. \label{eq_coefficent}
\end{equation}
We can also have the above relation from the viewpoint of the mapping~\cite{mapping}.

Using Eqs.~(\ref{eq_size_expansion}),~(\ref{eq_fixedpoint}) and
(\ref{eq_coefficent}), we can rewrite  Eq.~(\ref{eq_map}) as
\begin{eqnarray}
E_p(L_B, \rho) &=& E_p(L_A, \rho_c + Q^{y_t}(\rho-\rho_c))\nonumber \\
  &=& E_p(L_A, \rho_c + \varepsilon L_A^{-y_t})
\end{eqnarray}
with the scaling variable $\varepsilon \equiv
(\rho-\rho_c)L^{y_t}_B$. Since the system sizes $L_A$ and $L_B$
are arbitrary, we can choose $L_A$ as unity and rewrite $L_B$ as
$L$. Then we have
\begin{equation}
E_p(L,  \rho) = E_p(1,\rho_c + \varepsilon) \equiv F(\varepsilon),
\label{eq_finite_size_mapping}
\end{equation}
with the scaling function $F$.
Equation~(\ref{eq_finite_size_mapping}) implies that the
finite-size scaling theory corresponds to the first order
approximation of the expansion in Eq.~(\ref{eq_size_expansion}).

\begin{figure}[tbp]
\begin{center}
\includegraphics[width=.82\linewidth]{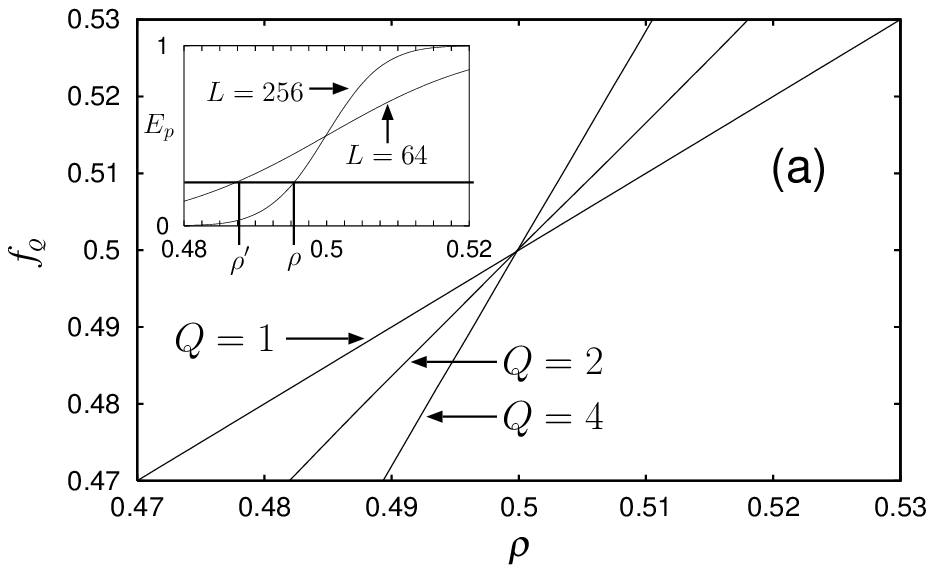}
\includegraphics[width=.82\linewidth]{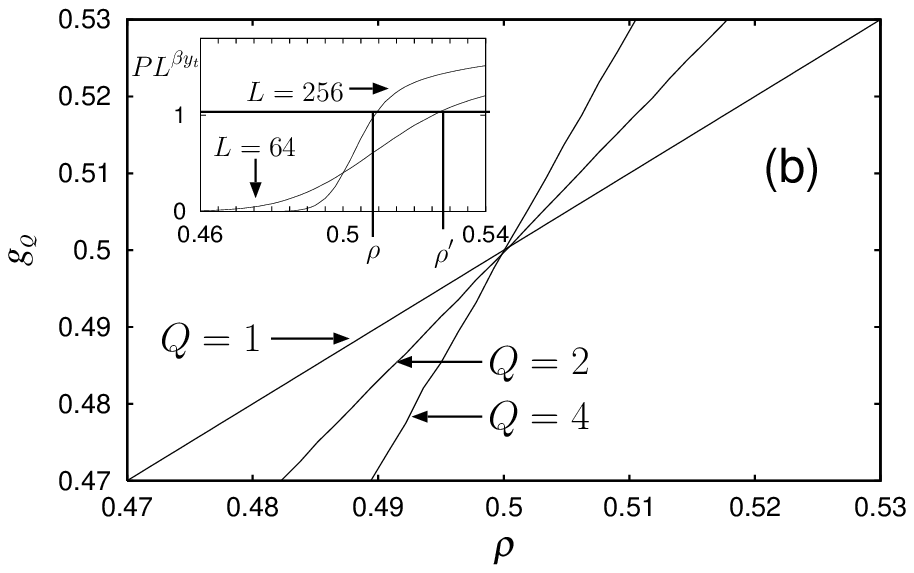}
\end{center}
\caption{ Mapping functions of the bond percolation model on
square lattices with free boundary conditions (a) $f_Q$ for the
existence probabilities $E_p$ as a function of $\rho$, and (b)
$g_Q$ for the percolation probabilities $P$ as a function of
$\rho$.  The insets in (a) and (b) show how to construct the
mapping functions for $E_p$ and $P$, respectively.}
\label{fig_mapping}
\end{figure}

To see the behavior of the mapping function $f_Q$, we determine
$f_Q$ from numerically obtained existence probabilities. The
procedures to determine $f_Q$ is as follows. (i)~Prepare graphs of
the existence probabilities vs.~density. (ii)~Draw a line parallel
to the horizontal axis. (iii)~Determine the two intersection point
$\rho$ and $\rho'$ (see the inset of Fig.~\ref{fig_mapping}(a)).
(iv)~Plot all the pairs $(\rho,\rho')$ by sweeping from $E_p = 0$
to $1$, then the function $\rho'= f_Q(\rho)$ is obtained. The
calculated mapping functions of the bond percolation on square
lattices are shown in Fig.~\ref{fig_mapping}(a) with aspect ratio
$R=1$ and system sizes $L = 64, 128$ and $256$. Free boundary
conditions are taken for this and other systems in the present
paper. All of these functions have linear forms and have the
single intersection point, namely the critical point. The ratio of
the slope of these lines gives the value of the critical exponent
$\nu=1/y_t$.

Similar arguments can be applied for the percolation probability
$P$. A mapping function $g_Q$ is defined by,
\begin{equation}
L_B^{\beta y_t} P(L_B,\rho,R) = L_A^{\beta y_t} P(L_A,g_Q(\rho),R)
\label{g}
\end{equation}
with a system size ratio $Q = L_B/L_A$. Here, we defined the
mapping function with the additional factors $L_B^{\beta y_t}$ and
$L_A^{\beta y_t}$ on the left-hand and right-hand sides of Eq.
(\ref{g}). With such factors, both sides of Eq. (\ref{g}) are
proportional to the FSSF for $P$ as can be seen from Eq.
(\ref{e0}). For $E_p$, the critical exponent is 0 \cite{Ep} and we
need not add such factors in Eq. (\ref{eq_map}) to define the
mapping function.
The functions
$g_Q$ are shown in Fig.~\ref{fig_mapping}(b). Figure 1 shows that
two kinds of functions $f_Q$ and $g_Q$ are identical near the
critical point.

\begin{figure}[tbp]
\begin{center}
\includegraphics[width=.82\linewidth]{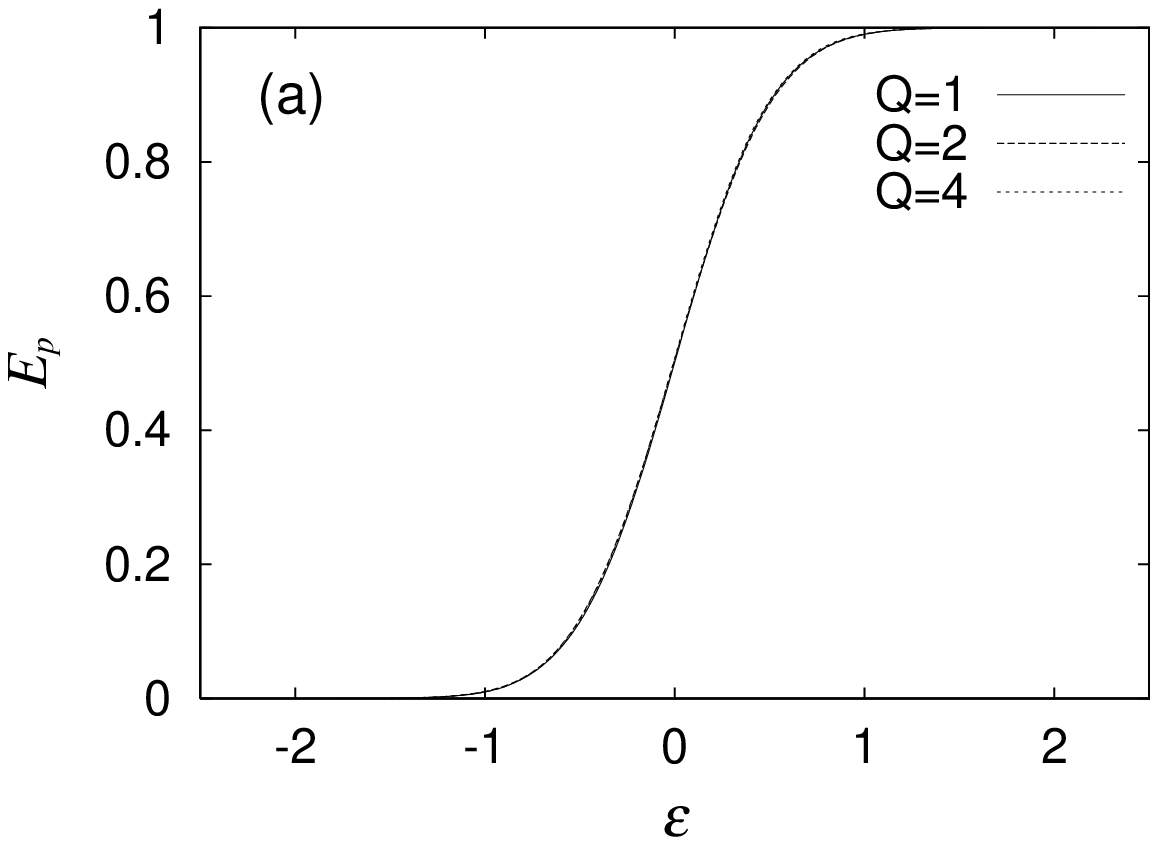}
\includegraphics[width=.82\linewidth]{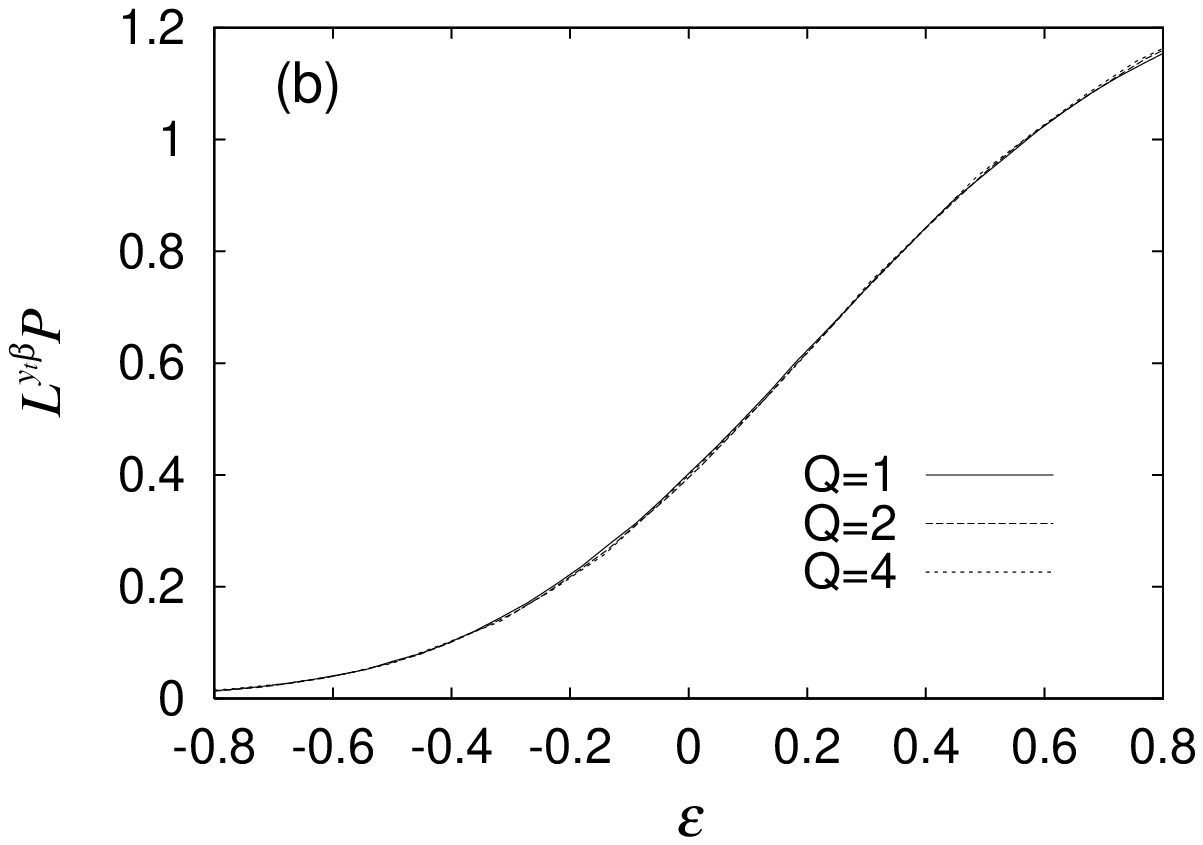}
\end{center}
\caption{ UFSSFs of (a) existence probabilities and (b) percolation probabilities
corresponding to Fig.~\ref{fig_mapping}.
The aspect ratio is $R=1$, and the system sizes are
$L=64 (Q=1)$, $128 (Q=2)$, and $256 (Q=4)$, respectively.
The scaling variable is denoted by $\varepsilon \equiv (\rho-\rho_c)L^{y_t}$.
}
\label{fig_ufssf}
\end{figure}

The UFSSFs corresponding to mapping functions in
Fig.~\ref{fig_mapping} are shown in Fig.~\ref{fig_ufssf}. The
goodness of the scaling in these figures come from the fact that
the mapping functions are well approximated by the linear function
when the systems have the identical aspect ratio. For the systems
with different aspect ratios, the mapping functions for $E_p$
becomes highly non-linear function as shown in Fig. 3(b) below.
Therefore, the scaling with low-order approximation does not show
good data collapse for large values of the scaling variables~(see
Fig. 4 in Ref.~\cite{04prl}).


\vskip 2 mm

\section{Mapping function and superscaling}

As shown in Fig. 3(a), we define the mapping function $f_R(\rho)$
to map the existence probability of the bond percolation model on
$RL \times L$ lattice, $E_p(L,\rho,R)$, into the existence
probability of the bond percolation model on $L \times L$ lattice,
$E_p(L,f_R(\rho),1)$, with following equation
\begin{equation}
E_p(L,\rho,R) = E_p(L,f_R(\rho),1).
\label{eq_rmapping}
\end{equation}
The curves of the existence probabilities for the same $L$ and
different $R$ do not intersect, since their values at the point
$\rho_c$  of the thermodynamic  system are different
~\cite{Cardy}. Therefore, we introduce the effective critical
point $\rho_c'$, which satisfies the equation $$f_R(\rho_c') =
\rho_c,$$ and consider the expansion of the mapping function $f_R$
at $\rho_c'$ as
\begin{eqnarray}
f_R(\rho) &=& f_R(\rho_c') + \sum_{n=1} \left. \frac{\diff^n
f_R}{\diff \rho^n}
\right|_{\rho_c'} \frac{(\rho-\rho_c')^n}{n!} \\
&=& \rho_c + \left. \frac{\diff f_R}{\diff \rho} \right|_{\rho_c'}(\rho-\rho_c')+
\left. \frac{\diff^2 f_R}{\diff \rho^2} \right|_{\rho_c'} \frac{(\rho-\rho_c')^2}{2}
 + \cdots. \label{eq_expansion}
\end{eqnarray}
The expansion coefficients depend only on $R$. With similar
arguments in Ref.~\cite{mapping}, the aspect ratio dependence of
the differential coefficient of the first order can be assumed to
be
\begin{equation}
\left. \frac{\diff f_R}{\diff \rho} \right|_{\rho_c'} \sim R^{a},
\label{eq_powerlaw}
\end{equation}
and taking into account up to the first order derivative of the
expansion, we obtain the approximated mapping function as,
$f_R(\rho) - \rho_c \propto R^a (\rho-\rho_c')$. From the
finite-size scaling,
\begin{equation}
E_p(L,f_R(\rho),1) \sim F\left( \left( f_R(\rho) - \rho_c  \right)L^{y_t} \right),
\label{eq_fscaling}
\end{equation}
and Eq.~(\ref{eq_rmapping}), we can derive the superscaling form to be,
\begin{eqnarray}
E_p(L,\rho,R) &=& E_p(L,f_R(\rho),1) \\
&\sim& F\left( \left( f_R(\rho) - \rho_c  \right)L^{y_t} \right) \\
&\sim& F((\rho-\rho_c')L^{y_t}R^a).
\end{eqnarray}

\begin{figure}[t]
\includegraphics[width=.8\linewidth]{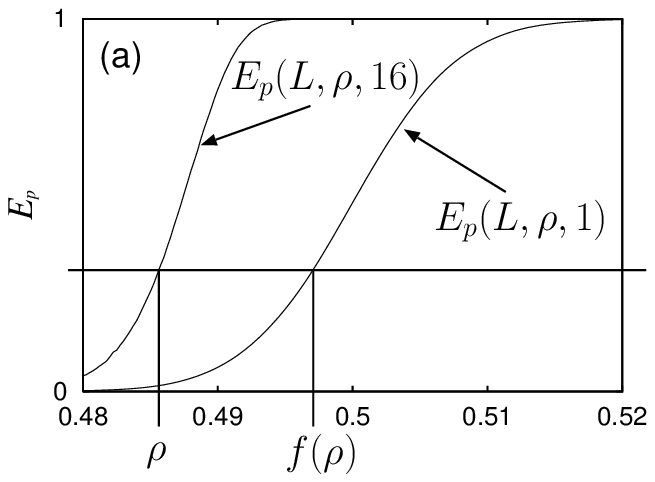}
\includegraphics[width=.8\linewidth]{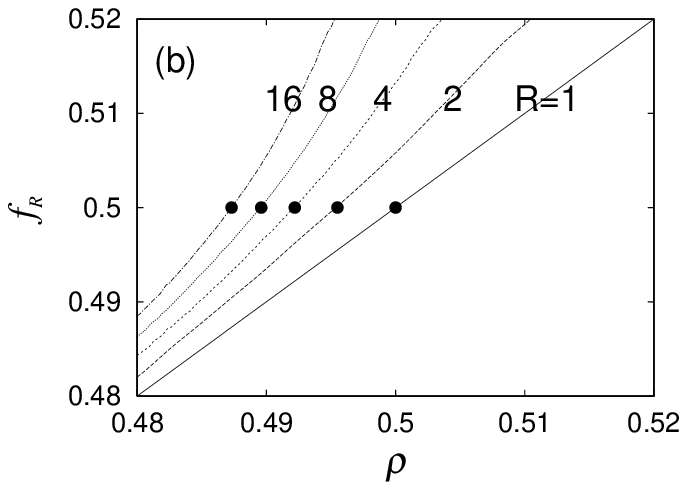}
\caption{ (a) Procedures to obtain the mapping functions of
existence probabilities with different aspect ratios. (b) Obtained
mapping functions of systems with $L=256$ and $R=1,2,4,8$ and $16$
(from bottom to top) .  The solid circles are the effective
critical points $\rho_c'$. } \label{fig_rmap_ep}
\end{figure}

The mapping functions $f_R$ of existence probabilities of the bond
percolation model on the square lattices are shown in
Fig.~\ref{fig_rmap_ep}(b) for the system size $L=256$ and aspect
ratios $R=1, 2, 4, 8$ and $16$. It shows that these functions do
not have any intersection points, since $E_p(L,\rho,R) >
E_p(L,\rho,1)$. Additionally, while the mapping functions of $f_Q$
in  Fig.~\ref{fig_mapping} have linear forms, the curves of $f_R$
in Fig. 3(b) for $R > 1$ are not linear.

\begin{figure}[t]
\includegraphics[width=.8\linewidth]{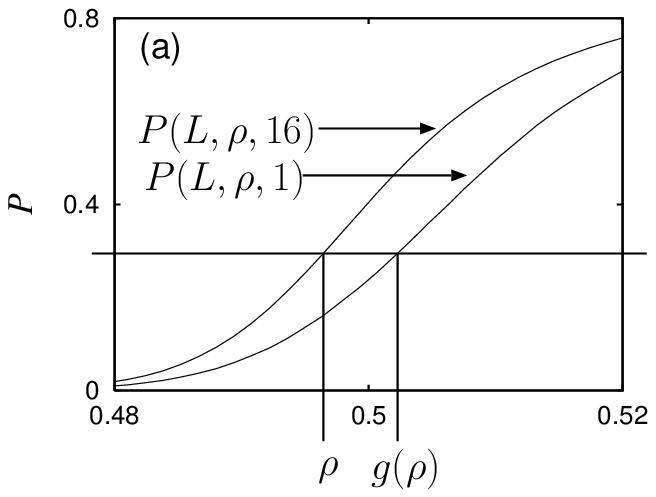}
\includegraphics[width=.8\linewidth]{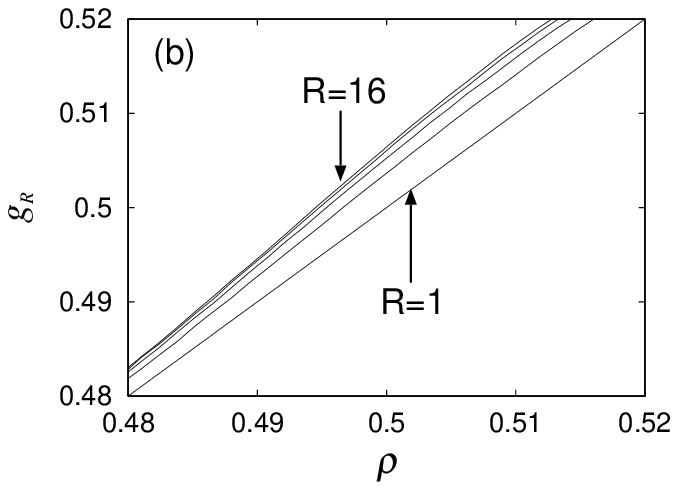}
\caption{ (a) Procedures to obtain the mapping functions of
percolation probabilities with different aspect ratios. (b)
Obtained mapping functions of systems with $L=128$ and $R=1,2,4,8$
and $16$ (from bottom to top). } \label{fig_rmap_P}
\end{figure}
\begin{figure}[htb]
\includegraphics[width=.8\linewidth]{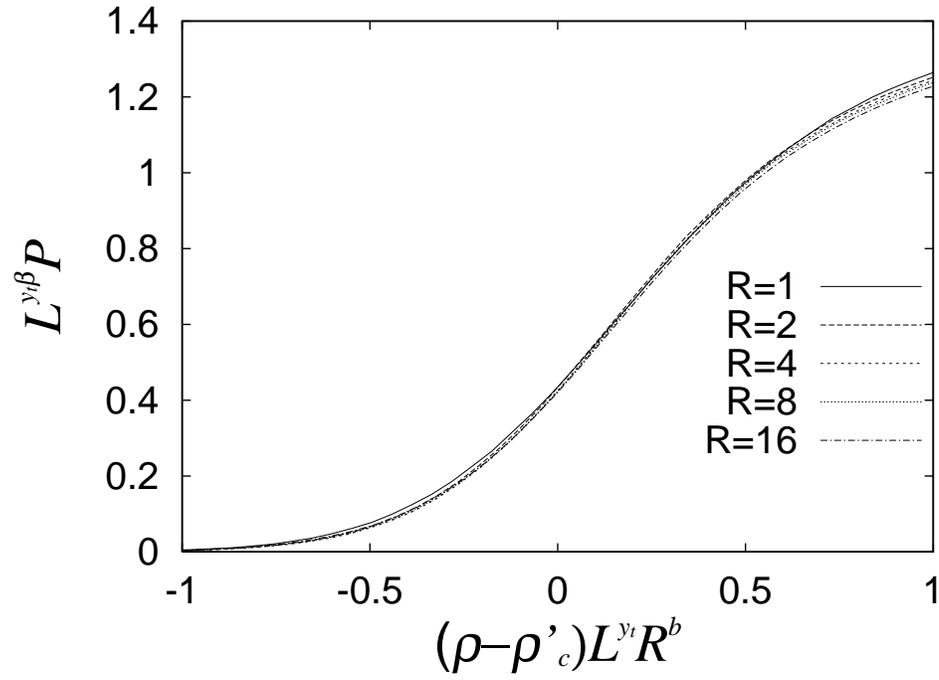}
\caption{ Superscaling plot of the percolation probabilities $P$
using Eq.~(\ref{eq_pscale}) for systems with $R = 1,2,4,8$ and
$16$, and $L=128$ with $b = 0.05$. } \label{fig_pscale}
\end{figure}

Consider the mapping of the percolation probabilities with
different aspect ratio as,
\begin{equation}
P(L,\rho,R) = P(L,g_R(\rho),1),
\label{eq_rmapping_p}
\end{equation}
with a mapping function $g_R$.
Similar to the case of the existence probability,
the behavior of the differential coefficient of $g_R$
can be assumed as
\begin{equation}
\left. \frac{\diff g_R}{\diff \rho} \right|_{\rho_c'} \sim R^b,
\end{equation}
with the effective critical point $\rho_c'$ defined by
$g_R(\rho_c') = \rho_c$ as shown in Fig. 4(a). Note that, the
value of effective critical point is different from the one
defined by $f_R(\rho_c') = \rho_c$. We can obtain the superscaling
formula as,
\begin{equation}
P(L,\rho,R) \sim L^{-\beta y_t} F\left( (\rho-\rho_c') L^{y_t} R^b \right).
\label{eq_pscale}
\end{equation}
This scaling form is different from the form
\begin{equation}
P(L,\rho,R) \sim (L^{y_t} R^b)^{-\beta} F\left( (\rho-\rho_c')
L^{y_t} R^b \right) \label{P04}
\end{equation}
obtained by a heuristic argument in \cite{04prl}. The mapping
functions of percolation probabilities are determined and shown in
Fig.~\ref{fig_rmap_P}(b). While they also do not have any
intersection points, they have almost linear forms. The scaling
plot based on Eq.~(\ref{eq_pscale}) is shown in
Fig.~\ref{fig_pscale}. It shows good scaling behavior especially
around at the effective critical point. Please note that the
right-hand sides of Eqs. (\ref{eq_pscale}) and (\ref{P04}) differ
only by the factor $R^{-b\beta}$ which is very close to 1.

\begin{figure}[htb]
\includegraphics[width=.8\linewidth]{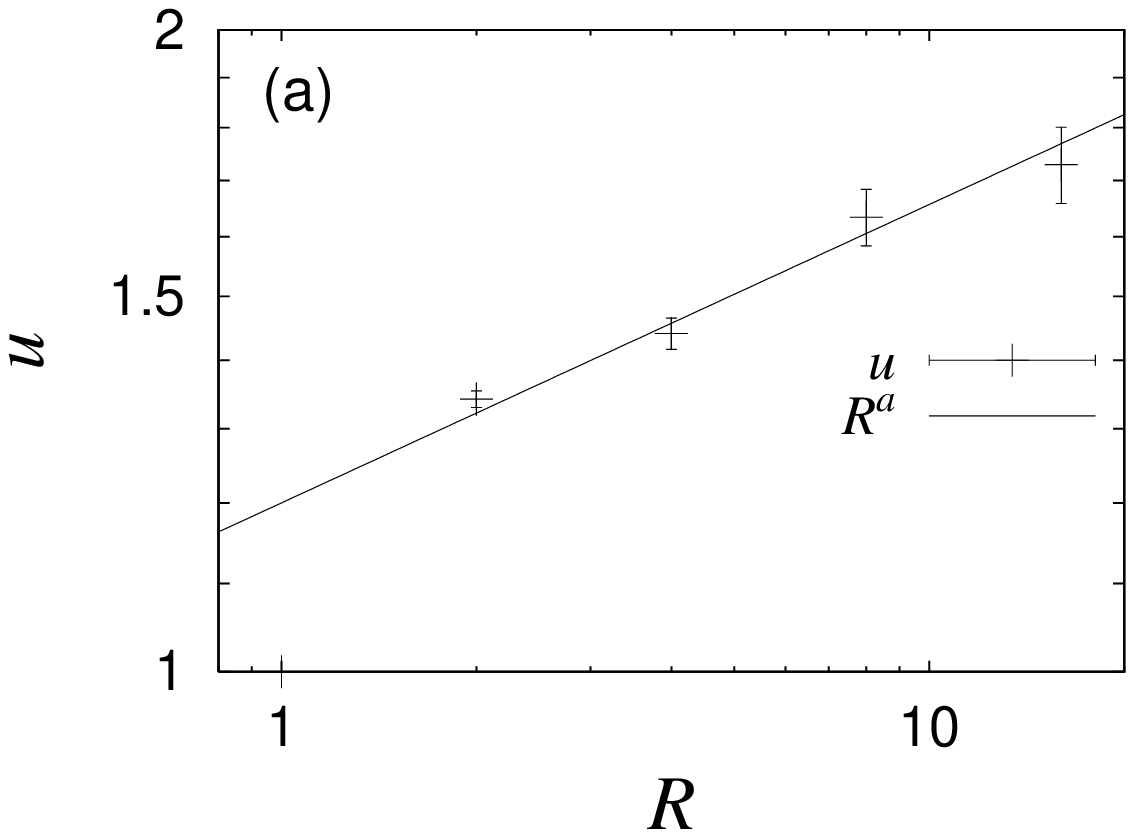}
\includegraphics[width=.8\linewidth]{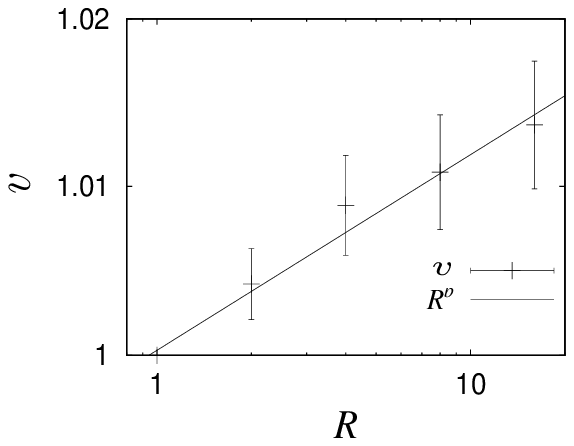}
\caption{  Plots of (a) $u = \diff f/\diff \rho|_{\rho_c'}$ and
(b) $v = \diff g/\diff \rho|_{\rho_c'}$ as a function $R$. The
solid lines are $R^{a}$ and $R^{b}$ with exponents $a = 0.14$ and
$b = 0.05$, respectively. Decimal logarithms are taken for both
axes.} \label{fig_slope}
\end{figure}

Since the mapping functions $f_R$ for $E_p$ have finite curvature,
the scaling formula $E_p(L,\rho,R) \sim F(\epsilon'L^{y_t}R^a)$
will be improved if we take into account higher order. This
nonlinear scaling form was shown in \cite{04prl}, and it
corresponds to approximate mapping function with a quadratic form
as $f_R(\rho) = c_2 \rho^2 + c_1 \rho + c_0$, which is the
expression obtained by taking into account up to the second order
derivatives in Eq.~(\ref{eq_expansion}). With the coefficients
$c_2, c_1$ and $c_0$ listed in \cite{04prl}, we calculate the
differential coefficients and show the results in
Fig.~\ref{fig_slope}(a). The mapping functions of the percolation
probabilities have almost linear form, we just fit them with
linear function to obtain the differentiations. The obtained
differential coefficients of the percolation probabilities are
shown in Fig.~\ref{fig_slope}(b). Figures \ref{fig_slope}(a) and
(b) show power law behavior $R^a$ and $R^b$ with values of
exponents $a = 0.14$ and $b = 0.05$ which are consistent with the
results in \cite{04prl}. Thus $a$ and $b$ can be obtained from
mapping functions $f_R$ and $g_R$ for $E_p$ and $P$, respectively.




\section{Summary and Discussion}

In summary, we have studied the aspect ratio dependence of FSSFs
of existence and percolation probabilities based on the idea of
the mapping functions. From the analysis on the numerically
obtained mapping functions, the superscaling behaviors are found
to be the first order approximation of the mapping functions at
the effective critical point. Since the mapping regime used here
is pretty general, we can apply it for any other systems and
geometric conditions.

Many lattice phase transition models, such as the Ising model, the
Potts model, the dilute Potts model, the hydrogen bonding model
for water, lattice hard-core particle models, etc, have been shown
to be corresponding to some correlated percolation models
\cite{corr-perco}. Some of such correlated percolation models have
been found to have good finite-size scaling behaviors
\cite{fsscp}. It is of interest to use mapping functions to study
super-scaling in such correlated percolation models and other
critical systems \cite{97pre}.

Another interesting development for critical systems is
finite-size corrections for lattice phase transition models
\cite{67jmp,ziff,99pre,Salas,fsc,07preTori}. Using exact partition
functions of the Ising model on finite lattices \cite{02jpaWu} and
exact finite-size corrections for the free energy, the internal
energy, and the specific heat of the Ising model \cite{fsc}, Wu,
Hu and Izmailian \cite{03preWu} obtained UFSSFs for the free
energy, the internal energy, and the specific heat with analytic
equations, which are free of simulation errors. Based on
finite-size corrections for the dimer model on finite lattices
\cite{03preDimer}, Izmailian, Oganesyan, Wu, and Hu
\cite{06preDimer} obtained UFSSFs for the dimer model with
analytic equations, which are also free of simulation errors. It
is of interest to extend above results for the Ising and dimer
models to rectangular domains with various aspect ratios. We can
then study superscaling behavior of physical quantities of the
Ising and dimer models with the help of mapping functions.


\vskip 5 mm

\centerline{\bf ACKNOWLEDGEMENTS}

\vskip 5 mm

This work was supported by National Science Council in Taiwan
under Grant No. 96-2911-M 001-003-MY3, National Center for
Theoretical Sciences in Taiwan, the Ministry of Education,
Science, Sports and Culture, Grant-in-Aid for Young Scientists
(B), 19740235 in Japan, and the 21st COE program of Nagoya
University. The computation was carried out using the facilities
of the Supercomputer Center, Institute for Solid State Physics,
University of Tokyo.

%
%

\begin{thebibliography}{99}
\bibitem{Ep} In the limit $L \to \infty$, $E_p=0$ for the bond
density $\rho$ smaller than $\rho_c$ and $E_p=1$ for bond density
$\rho$ larger than $\rho_c$. If we write $E_p\sim (\rho-\rho_c)^c$
for $\rho > \rho_c$, then the critical exponent $c$ is 0.
\bibitem{mapping}
The identity $f_{Q^{-1}}(f_Q(\rho)) = \rho$ directly leads to the
relation $f_{Q^{-1}}'|_{\rho_c} \times f_Q'|_{\rho_c}  = 1$ (
$f_Q'$ representing the derivative of $f_Q$ with respect to
$\rho$), because $f_Q(\rho_c)=\rho_c$ regardless of the value of
$Q$. Therefore the $Q$-dependence of the coefficient should have
the form $f_Q'|_{\rho_c} = Q^\alpha$.
\end{thebibliography}
\end{document}